\documentclass[12pt]{article}
\usepackage{amsfonts}
\usepackage{amssymb}
\usepackage{amsbsy}
\usepackage{graphics}

\input epsf.sty

\newlength{\TZ}
\newcommand{\BEQ}{\begin{equation}}     
\newcommand{\BEA}{\begin{eqnarray}}
\newcommand{\EEQ}{\end{equation}}       
\newcommand{\EEA}{\end{eqnarray}}
\def\be{\begin{equation}}
\def\ee{\end{equation}}
\def\ba{\begin{eqnarray}}
\def\ea{\end{eqnarray}}
\newcommand{\eps}{\varepsilon}          
\newcommand{\D}{{\rm d}}                
\newcommand{\wit}[1]{\widetilde{#1}}    

\renewcommand{\vec}[1]{{\boldsymbol{#1}}} 

\newcommand{\zeile}[1]{\vskip #1 \baselineskip} 

\catcode`\@=11
\def\numberbysection{\@addtoreset{equation}{section}
        \def\theequation{\thesection.\arabic{equation}}}
\numberbysection

\begin{document}

\begin{titlepage}

~~~ 

\vskip 1.5 cm
\begin{center}
{\Large \bf On lower bounds of the solutions of some simple 
reaction-diffusion equations}
\end{center}

\vskip 2.0 cm
\centerline{{\bf Malte Henkel}$^{a,b}$}
\vskip 0.5 cm
\centerline {$^a$Laboratoire de Physique des
Mat\'eriaux,\footnote{Laboratoire associ\'e au CNRS UMR 7556}
Universit\'e Henri Poincar\'e Nancy I,}
\centerline{ B.P. 239,
F -- 54506 Vand{\oe}uvre l\`es Nancy Cedex, France\footnote{permanent address}}
\vskip 0.5 cm
\centerline {$^{b}$Centro de F\'{\i}sica Te\'orica e Computacional,
Universidade de Lisboa,}
\centerline{ Av. Prof. Gama Pinto 2, P--1649-003 Lisboa, Portugal}

\begin{abstract}
The mean-field reaction-diffusion equations of the diffusive 
pair-an\-ni\-hi\-la\-tion and triplett-an\-ni\-hi\-la\-tion processes are 
considered. A direct lower bound on the time-dependent mean particle-density 
is derived. The results are applied to
the mean-field theory of the diffusive pair-contact process.  
\end{abstract}

\zeile{4}
\end{titlepage}

\section{Introduction}

The interplay of reaction kinetics with particle diffusion has since a long 
time been a topic of intensive research. Here we shall consider the following
reaction-diffusion equations
\BEQ \label{1:gl:1}
\partial_t a(t,\vec{r}) = \Delta a(t,\vec{r}) - \lambda a(t,\vec{r})^2
\EEQ
and
\BEQ \label{1:gl:1a}
\partial_t b(t,\vec{r}) = \Delta b(t,\vec{r}) - \mu b(t,\vec{r})^3
\EEQ
where $t\in \mathbb{R}_+$ and $\vec{r}\in\mathbb{R}^n$ are time and space
coordinates, $\Delta$ is the Laplacian and $\lambda,\mu>0$ are constant 
reaction rates. It is well-known that eq.~(\ref{1:gl:1}) provides a mean-field
description of the pair-annihilation process $A+A\to\emptyset$ together with
single-particle diffusion and similarly, (\ref{1:gl:1a}) describes triplett
annihilation $A+A+A\to\emptyset$ (we have rescaled the diffusion-constant to 
one). Still, the derivation of the long-time behaviour of solutions of such
non-linear partial differential equations is not completely trivial. It is 
convenient to consider eq.~(\ref{1:gl:1}) inside a spatial domain 
$\Omega\subset\mathbb{R}^n$ and to define the mean densities
\BEQ \label{1:gl:2}
\overline{a} = \overline{a}(t) = 
\frac{1}{|\Omega|} \int_{\Omega} \!\D \vec{r}\: a(t,\vec{r}) \;\; ; \;\;
\overline{b} = \overline{b}(t) = 
\frac{1}{|\Omega|} \int_{\Omega} \!\D \vec{r}\: b(t,\vec{r})
\EEQ
where $|\Omega|$ denotes the volume of $\Omega$. 

The most simplistic treatment of (\ref{1:gl:1}) 
simply suppresses the diffusion term, which
leads to $\partial_t\overline{a}_s = -\lambda \overline{a}_s^2$ and the 
solution $\overline{a}_s(t)=a_0/(1+a_0 \lambda t)$. Then, it may be asked
to what extent this drastic simplification may be justified. This, and more
generally the long-time behaviour of the space-time-dependent particle
density $a(t,\vec{r})$, has received a lot of attention. For example, 
{\it a priori} estimates such as the 
strong maximum principle may be invoked to obtain bounds 
$v_{-}(t)\leq a(t,\vec{r})\leq v_{+}(t)$ where $v_{\pm}(t)$ satisfy
the simplistic equation $\partial_t v_{\pm} = -\lambda v_{\pm}^2$ with the
initial conditions $v_{-}(0)\leq a(0,\vec{r})\leq v_{+}(0)$ 
\cite[p. 94]{Smol83}. The latter conditions might however be difficult to meet
for very `rough' initial data $a(0,\vec{r})$. 
For domains with a finite (and `small') volume $|\Omega|$, one may define
an invariant region $\Sigma$ and define $\sigma:=\eta-M$, where
$\eta$ is the principal eigenvalue of $-\Delta$ on $\Omega$ and 
$M=\max_{\Sigma} |\nabla \lambda a^2|$. If $\sigma>0$, then it can be shown 
that $a(t,\vec{r})$ converges exponentially fast with a characteristic time 
$1/\sigma$ towards the solution $\overline{a}_s(t)$ \cite[p. 223]{Smol83}.
However, the implied exponential approach need not hold any longer in spatially
infinite regions. Then methods based on a scaling {\em ansatz} of the form
$a(t,\vec{r})=t^{-\alpha/2} f(\vec{r}t^{-1/2})$ permit to extract the
long-time asymptotics of the solution from phenomenological scaling
\cite{Corn93} or by using rigorous renormalization-group 
arguments \cite{Bric94}. This kind of argument can also be extended to
systems of reaction-diffusion equations and in particular 
the reactions fronts in the two-component system $kA+kB\to\emptyset$ with
initial conditions such that there is a reaction front between A-rich and 
B-rich regions. In $n=1$ dimension, it was shown rigorously for 
$k=1$ \cite{Sche93} and $k\geq 4$ \cite{Baal00} that the particle densities
and the reaction front satisfy a multiscaling behaviour and that 
the convergence towards the scaling solutions is controlled by algebraically 
(and not exponentially) small corrections in $t$. On the diffusive scale, when
$|\vec{r}|/\sqrt{t}\gg 1$, the problem essentially reduces to 
$\partial_t a = \Delta a - a^k$ \cite{Baal00}. Reaction-diffusion systems
of the form $\partial_t \vec{a}=
\Delta\vec{a}-\nu\vec{a}-\vec{f}(\vec{a})+\vec{g}$ with 
$\vec{a}|_{t=0}=\vec{a}_{(0)}$, $\vec{a}|_{\partial\Omega}=\vec{0}$, 
$\nu>0$ and suitable assumptions on $\vec{f}$ are reviewed in \cite{Zeli03}. 
The long-time behaviour of the unique solution (roughly, on the Sobolev space 
$W^{2,q}(\Omega)$ where $\vec{g}\in L^q(\Omega)$ with $q>\max(2,n/2)$) is
described in terms of an attractor whose complexity can be analysed through 
its Kolmogorov entropy in great detail. 

Here we wish to present a simple direct estimate on the mean 
densities $\overline{a}(t)$ and $\overline{b}(t)$. 
We have (see section 2 for notations) \\

\noindent
{\bf Theorem:} {\it (i) Let $a\in C^1(\mathbb{R}_+;W^{2,2}(\Omega))$ 
be an (almost) everywhere non-negative solution of (\ref{1:gl:1}). 
Let in addition be
$\vec{\nabla} a=0$ on the boundary $\partial\Omega$. Then there is a positive 
and $|\Omega|$-independent constant $\lambda'$ such that the the mean density 
$\overline{a}$ satisfies}
\BEQ \label{1:gl:3}
-\lambda'\,\overline{a}(t)^2 \leq \partial_t \overline{a}(t) 
\leq -\lambda \overline{a}(t)^2 
\EEQ
{\it (ii) If $b\in C^1(\mathbb{R}_+;W^{2,3}(\Omega))$ is 
an (almost) everywhere non-negative solution of (\ref{1:gl:1a}) such that
$\left.\vec{\nabla} b\right|_{\partial\Omega}=0$, 
there is a positive and $|\Omega|$-independent constant $\mu'$ such that}
\BEQ \label{1:gl:3a}
-\mu'\,\overline{b}(t)^3 \leq \partial_t \overline{b}(t) 
\leq -\mu \overline{b}(t)^3 
\EEQ
{\it For $n\leq 3$, the condition $b\in 
C^1(\mathbb{R}_+;W^{2,2}(\Omega)\cap L^3(\Omega))$ is sufficient.}

The Sobolev space $W^{2,p}(\Omega)$ may be too restrictive for unbounded
domains. In section 2 we define a generalized space $\wit{W}^{2,p}(\Omega)$ for
which the theorem still holds and which includes spatially homogeneous 
initial states. 

The upper bound in (\ref{1:gl:3}) has been known since a long 
time \cite{Avra90}. 
In particular, (\ref{1:gl:3},\ref{1:gl:3a}) imply for $t$ sufficiently large
(i.e. $a_0\lambda' t>1$ and $2 b_0^2\mu't>1$, respectively, where
$a_0>0$ and $b_0>0$ are the initial mean densities) the bounds 
\BEQ \label{1:gl:4} 
\frac{1}{2\lambda'} \leq t\, \overline{a}(t) \leq \frac{1}{\lambda} 
\;\; ; \;\;
\frac{1}{2\sqrt{\mu'}} \leq t^{1/2}\,\overline{b}(t) \leq\frac{1}{\sqrt{2\mu}}
\EEQ
It is admissible to take the infinite-volume limit $|\Omega|\to\infty$. 
Eq.~(\ref{1:gl:4}) is in full agreement with the results established earlier 
by different means \cite{Baal00,Bric94,Corn93,Gmir84,Smol83}. 
For similar upper bounds on $||a||_p$, $||b||_p$ (with $1\leq p\leq \infty$) 
which explicitly depend on the initial data, see \cite{Gmir84}. We point out 
that our derivation makes no explicit reference to the initial conditions 
(beyond the requirement $a(t,\vec{r})\geq 0$, $b(t,\vec{r})\geq 0$) and neither
a scaling ansatz is needed. The bounds (\ref{1:gl:4}) reproduce the expected 
mean-field scaling $\overline{a}(t)\sim t^{-1}$. For the mathematically 
oriented reader we recall that in low dimensions $n<2$, the description of 
the diffusive pair-annihilation processes through a more microscopic approach 
such as a master equation (where fluctuations are taken into account) leads to 
a different long-time behaviour $\overline{a}_{\rm micro}(t)\sim t^{-n/2}$, 
which has also been observed experimentally for $n=1$, see \cite{Henk03} for a 
recent review. This manifests once again the
character of equations such as (\ref{1:gl:1}) as mean-field approximations.

The approach of $a(t,\vec{r})$ towards the mean density can be described as
follows.

\noindent
{\bf Corollary:} {\it Under the same conditions as in the theorem, there is
a constant $K'\leq 1$ such that for times satisfying the conditions used
in eq.~(\ref{1:gl:4})}
\BEA
\frac{1}{|\Omega|} \int_{\Omega} \!\D\vec{r}\: 
\left(a(t,\vec{r}) - \overline{a}(t)\right)^2 
\leq  \frac{K'}{\lambda^2}\cdot t^{-2}
\nonumber \\
\frac{1}{|\Omega|} \int_{\Omega} \!\D\vec{r}\: 
\left(b(t,\vec{r}) - \overline{b}(t)\right)^2
\leq \frac{K'}{2\mu}\cdot t^{-1} 
\EEA
\zeile{1}

As an application, we consider the {\it pair-contact process} $2A\to\emptyset$, 
$2A\to 3A$ with single-particle diffusion ({\sc pcpd}). We consider a domain
$\Omega\subset\mathbb{R}^n$ with the boundary condition 
$\vec{\nabla} a|_{\partial\Omega}=0$. The mean-field reaction-diffusion equation is
\BEQ \label{1:gl:8}
\partial_t a(t,\vec{r}) = D \Delta a(t,\vec{r}) +\lambda a(t,\vec{r})^2 
-\mu a(t,\vec{r})^3
\EEQ
with constants $\lambda\in\mathbb{R}$ and $\mu>0$. 
If the diffusion constant $D=0$,
$a=a(t)$ evolves for $\lambda>0$ towards a steady-state density
$a_{\infty}=\lambda/\mu$, while $a_{\infty}=0$ for $\lambda\leq 0$. It is
known that \cite{Carl01}
\BEQ \label{1:gl:9}
a(t) - a_{\infty} \sim \left\{ \begin{array}{ll}
{\rm O}\left(e^{-t/\tau}\right) & \mbox{\rm ~~; if $\lambda >0$} \\
t^{-1/2} & \mbox{\rm ~~; if $\lambda =0$} \\
t^{-1} & \mbox{\rm ~~; if $\lambda <0$}
\end{array} \right.
\EEQ
as $t\to\infty$ and where $\tau>0$ is a known constant. On the other hand, 
for a non-vanishing diffusion constant,
we scale to $D=1$ and have for the mean density
\BEQ \label{1:gl:10}
\lambda \overline{a}(t)^2 - \mu' \overline{a}(t)^3 
\leq \partial_t \overline{a}(t) 
\leq \lambda' \overline{a}(t)^2 - \mu \overline{a}(t)^3
\EEQ
for $\lambda\geq 0$ and
\BEQ\label{1:gl:11}
-|\lambda'|\, \overline{a}(t)^2 - \mu' \overline{a}(t)^3 
\leq \partial_t \overline{a}(t) 
\leq -|\lambda|\, \overline{a}(t)^2 - \mu \overline{a}(t)^3
\EEQ
for $\lambda\leq 0$, respectively and we can now let $|\Omega|\to\infty$, if
we so desire. For $\lambda>0$, there is an active 
steady-state with density $\lambda/\mu' \leq a_{\infty}\leq \lambda'/\mu$
but if $\lambda\leq 0$, one has $a_{\infty}=0$. Furthermore, eq.~(\ref{1:gl:9}) 
can be taken over. This proves the existence of a continuous steady-state
transition at $\lambda_c=0$ of the mean-field equation (\ref{1:gl:8}). 
Finally, the approach of $a(t,\vec{r})$ towards the mean density is
according to
\BEQ \label{1:gl:12}
\frac{1}{|\Omega|} \int_{\Omega} \!\D\vec{r}\: 
\left(a(t,\vec{r}) - \overline{a}(t)\right)^2 
\lesssim \left\{ \begin{array}{ll}
{\rm O}\left(e^{-2t/\tau'}\right) & \mbox{\rm ~~; if $\lambda >0$} \\
t^{-1} & \mbox{\rm ~~; if $\lambda =0$} \\
t^{-2} & \mbox{\rm ~~; if $\lambda <0$}
\end{array} \right.
\EEQ
In $n=1$ dimension, fluctuation effects create a very rich behaviour of the
{\sc pcpd} which is under active investigation, see \cite{Henk04} for a review. 

In section 2, we recall some inequalities which are needed in the proofs.
The upper bounds in (\ref{1:gl:3},\ref{1:gl:3a}) are derived in section 3 
and in section 4, the lower bounds are obtained.

\section{Mathematical background}

We recall here some standard notations and some inequalities which will
be needed in establishing the lower bound in (\ref{1:gl:3},\ref{1:gl:3a}). 
We shall work with the $p$-norms, for $1\leq p < \infty$ 
\BEQ
|| u ||_p := \left( \int_{\Omega} \!\D\vec{r}\: |u(\vec{r})|^p\right)^{1/p}
\EEQ
Denote by $L^p(\Omega)$ the space of (equivalence classes of) functions with
$||u||_p$ finite. Here and in the following $\Omega\subset \mathbb{R}^n$. 
Furthermore, if $\Omega$ has a boundary, it is assumed to be sufficiently 
smooth throughout. The space $L^{\infty}(\Omega)$ is defined with respect
to the supremum norm $||u||_{\infty}=\mbox{\rm ess sup}_{\Omega} |u(\vec{r})|$. 
For derivatives, we use the multiindex notation 
$\alpha=(\alpha_1,\ldots,\alpha_n)\in\mathbb{N}_0^n$ and where
$|\alpha|:=\alpha_1+\ldots+\alpha_n$. Then derivatives are denoted by
\BEQ
\partial^{\alpha} u = \frac{\partial^{|\alpha|}u}{(\partial r_1)^{\alpha_1}
\ldots (\partial r_n)^{\alpha_n}}
\EEQ
These derivatives can be taken to be weak (distributional) derivatives. 
The Sobolev space is
\BEQ
W^{k,p}(\Omega) := \left\{ u\in L^p(\Omega), \partial^{\alpha} u\in L^p(\Omega)
\mbox{\rm ~~for all $|\alpha|\leq k$}\right\}
\EEQ
with its norm $||u||_{k,p} = \sum_{|\alpha|\leq k} ||\partial^{\alpha} u||_p$. 
Finally, $C^1(\mathbb{R}_+;W^{k,p}(\Omega))$ is the space of functions which
are continuously differentiable with respect to $t$ for all times 
$0\leq t<\infty$ and whose values at any given $t$ are in $W^{k,p}(\Omega)$. 

\newpage \typeout{** hier ist ein Seitenvorschub !! **}

After these preparations, we can state some known results which we need later.
The first one is the Gagliardo-Nirenberg inequality, 
see \cite{Frie69,Shat00}. \\

\noindent
{\bf Lemma 1.} {\it For functions $u\in W^{k,p}(\Omega){\cap} L^q(\Omega)$ with
$1\leq q\leq\infty$ and for any integer $0\leq j < k$, there is a constant $C>0$
such that}
\BEQ
||\partial^j u||_r \leq C\, ||u||_q^{1-\theta}\, ||\partial^k u||_p^{\theta}
\EEQ
{\it where}
\BEQ
\frac{1}{r} - \frac{j}{n} = \frac{1-\theta}{q} +
\theta\left(\frac{1}{p}-\frac{k}{n}\right)
\EEQ
{\it and if $1\leq p<n/(k-j)$, then $j/k\leq\theta\leq 1$. On the other hand,
if $1\leq p =n/(k-j)$, only $j/k\leq \theta<1$ is allowed.}

We shall need two special cases of this. For $n\geq 2$, we set $p=r=2$, $k=q=1$.
Then $j=0$ and $\theta=n/(2+n) \in[\frac{1}{2},1)$. We get (here and in the 
following, we suppress the integration variable and write 
$\int_{\Omega} = \int_{\Omega} \!\D\vec{r}$) 
\BEQ \label{2:gl:GN1}
\int_{\Omega} |u|^2 \leq C_{1} \left( \int_{\Omega} |u| \right)^{4/(2+n)}
\left( \int_{\Omega} |\partial u|^2\right)^{n/(2+n)}
\EEQ
Eq.~(\ref{2:gl:GN1}) does not hold if $n<2$. 
For $1\leq n\leq 2$, we set $r=2$, $p=q=k=1$. Then $j=0$, 
$\theta=n/2\in[\frac{1}{2},1]$ and
\BEQ \label{2:gl:GN2}
\int_{\Omega} |u|^2 \leq C_{2} \left( \int_{\Omega} |u| \right)^{2-n}
\left( \int_{\Omega} |\partial u|\right)^{n}
\EEQ
where the constants $C_{1,2}$ equal $C^2$ from Lemma 1. It can be checked 
from dimensional analysis that $C_{1}$ and $C_{2}$ are independent of 
$|\Omega|$. 

Next, we quote an inequality due to Nirenberg, see \cite{Frie69}.\\

\noindent
{\bf Lemma 2.} {\it For $u\in W^{2,p}(\Omega)$, $p\geq 1$,
there exists a constant $\eps_0=\eps_0(p,\Omega)$ such that for any
$\eps$ with $0<\eps<\eps_0$, there is a positive constant $c=c(p,\Omega)$ 
such that}
\BEQ \label{2:gl:N}
\int_{\Omega} \!\D\vec{r}\: |\partial u|^p \leq \frac{c}{\eps} 
\int_{\Omega} \!\D\vec{r}\: |u|^p + 
\eps \sum_{|\alpha|=2} \int_{\Omega} \!\D\vec{r}\: |\partial^{\alpha} u|^p
\EEQ

Finally, we quote Poincar\'e's inequality, see \cite{Shat00}. Let $B_R(0)$ be
the ball of radius $R$ around the origin. \\

\noindent
{\bf Lemma 3.} {\it For any $u\in W^{1,p}(B_R(0))$ with $1<p<\infty$, there
is a positive constant $C_{\rm P}^{(p)}$ such that}
\BEQ \label{2:gl:P}
\int_{B_R(0)} \!\D \vec{r}\: |u - \overline{u}|^p \leq C_{\rm P}^{(p)} 
|B_{R}(0)|^{p/n} \int_{B_R(0)} \!\D \vec{r}\: |\partial u|^p 
\EEQ
{\it and the mean value $\overline{u}$ is defined in analogy with
(\ref{1:gl:2}).}

It is sometimes desirable to consider spaces which are less restrictive than 
the spaces $L^{p}(\Omega)$. We define $\wit{L}^p(\Omega)$ as the space of 
(equivalence classes of) functions such that 
$m_p(u):=|\Omega|^{-1} ||u||_p^p$ is finite. For unbounded domains
(e.g. $\Omega=\mathbb{R}^n$) a limit procedure must be used in the definition
of $m_p(u)$. We also set 
\BEQ
\wit{W}^{k,p}(\Omega) := \left\{ u\in \wit{L}^p(\Omega), 
\partial^{\alpha} u\in \wit{L}^p(\Omega)
\mbox{\rm ~~for all $|\alpha|\leq k$}\right\}
\EEQ
As an example, consider the function $f:\mathbb{R}\to\mathbb{R}$,
$x\mapsto f(x) =f_0\ne 0$. While $f\in\wit{L}^p(\mathbb{R})$, since
$m_p(f)=|f_0|^p$, clearly $f\not\in L^p(\mathbb{R})$. Lemmas 2 and 3
readily extend to the space $\wit{W}^{k,p}(\Omega)$. 

\section{The upper bound}

We briefly recall the proof of the upper bound in (\ref{1:gl:3}), following
\cite{Avra90}. The mean density satisfies
\BEA
\partial_t \overline{a} &=& \frac{1}{|\Omega|} \int_{\Omega}
\!\D\vec{r}\: \Delta a - 
\frac{\lambda}{|\Omega|}\int_{\Omega} \!\D\vec{r}\: a^2
\nonumber \\
&=& \frac{1}{|\Omega|} \int_{\partial\Omega} \!\D\vec{\sigma}\cdot \vec{\nabla}
a - \frac{\lambda}{|\Omega|}\int_{\Omega} \!\D\vec{r}\: a^2
\EEA
where $\vec{\sigma}$ is a normal vector to the boundary $\partial\Omega$. 
The first term describes the flux of particles through the boundary and 
vanishes either in the limit of large volumes $|\Omega|\to\infty$ or else if
the boundary condition $\vec{\nabla}a|_{\partial\Omega}=0$ is imposed. 
Then $\partial_t\overline{a}=-\lambda\overline{a^2}\leq 
-\lambda\overline{a}^{\,2}$ by the Cauchy-Schwarz inequality. 

A similar argument works for triplett annihilation. By H\"older's inequality,
$\partial_t\overline{b}=-\mu\overline{b^3}\leq -\mu\overline{b}^{\,3}$. 

\section{The lower bound}

In order to obtain the lower bound in (\ref{1:gl:3}), 
we recall from section 3 that
$\partial_t \overline{a} = - \lambda\overline{a^2}$. The right-hand side
is now estimated through the Gagliardo-Nirenberg inequality. We have to
distinguish the cases $n\geq 2$ and $n\leq 2$ and obtain from 
eqs.~(\ref{2:gl:GN1}) and (\ref{2:gl:GN2})
\BEQ \label{4:gl:1}
\partial_t \overline{a} \geq 
\left\{ \begin{array}{ll} 
-\lambda C_1 |\Omega|^{-1}\,\left(\int_{\Omega} a\right)^{4/(2+n)} 
\left( \int_{\Omega}|\nabla a|^2\right)^{n/(2+n)} & 
\mbox{\rm ~~; if $n\geq 2$} \\
-\lambda C_2 |\Omega|^{-1}\, \left(\int_{\Omega} a\right)^{\,2-n}
\left( \int_{\Omega}|\nabla a|^2\right)^{n/2} & 
\mbox{\rm ~~; if $n\leq 2$} 
\end{array} \right.
\EEQ
where for $n\leq 2$ the Cauchy-Schwarz inequality was used again. 
Next, we need an upper estimate for $\int_{\Omega}|\nabla a|^2$, which is
provided by the following 

\noindent
{\bf Proposition:} {\it For $u\in W^{2,p}(\Omega)$ there is a constant $c>0$ and
an $\eps_*$ such that $0<\eps_*<\infty$ and that for all $\eps<\eps_*$ one has}
\BEQ
\int_{\Omega} |\nabla u|^p \leq \frac{2c}{\eps} \int_{\Omega} |u|^p
\EEQ

\noindent 
{\bf Proof:} If $\sum_{|\alpha|=2} \int_{\Omega} |\partial^{\alpha} u|^p =0$, 
the proposition holds true trivially, because of (\ref{2:gl:N}). We can thus 
suppose that $\sum_{|\alpha|=2}\int_{\Omega} |\partial^{\alpha} u|^p >0$. 
Next, the function $f(x) := A/x + B x$, where $A,B$ are positive 
constants, has an absolute minimum at $x_*=\sqrt{A/B}$. For $x<x_*$, the
first term dominates over the second and $f(x)<2A/x$ for all $x<x_*$. We apply
this to the inequality (\ref{2:gl:N}) of Lemma 2. The right-hand side is 
minimal if $\eps=\eps_*$, where
\begin{displaymath}
\eps_* = \min \left( 
\left( \frac{c \int_{\Omega}|u|^p}{\sum_{|\alpha|=2}
\int_{\Omega}|\partial^{\alpha} u|^p}\right)^{1/2}, \eps_0 \right)
\end{displaymath}
with the $\eps_0$ of Lemma 2. Then the assertion follows. \hfill q.e.d.

The extension to $\wit{W}^{2,p}(\Omega)$ is immediate. 

Therefore, setting $p=2$ and appealing to dimensional analysis, 
there is a positive constant $K>0$ such that 
\newpage \typeout{** hier ist ein Seitenvorschub !! **}
\BEA
\int_{\Omega} |\nabla a|^2 &\leq& K |\Omega|^{-2/n} \int_{\Omega} a^2
\nonumber \\
&=& K |\Omega|^{-2/n} \int_{\Omega} \left[ (a-\overline{a})^2 +
2\overline{a}\,(a-\overline{a}) + \overline{a}^{\,2} \right]
\nonumber \\
&=& K |\Omega|^{1-2/n}\overline{a}^{\,2} +  
K |\Omega|^{-2/n} \int_{\Omega} (a-\overline{a})^2
\nonumber \\
&\leq& 2K |\Omega|^{1-2/n}\overline{a}^{\,2}
\label{4:gl:4}
\EEA
{}From the eqs.~(\ref{4:gl:1}) it follows 
\BEQ
\partial_t\overline{a} \geq 
\left\{ \begin{array}{ll} 
-\lambda C_1 (2K)^{n/(2+n)} \: \overline{a}^{\,2} & \mbox{\rm ~~; if $n\geq 2$} 
\\
-\lambda C_2 (2K)^{n/2} \: \overline{a}^{\,2}     & \mbox{\rm ~~; if $n\leq 2$} 
\end{array} \right.
\EEQ
This is exactly the form asserted in the theorem and we can identify the
effective reaction rate
\BEQ
\lambda' := \left\{ \begin{array}{ll}
\lambda C_1 (2K)^{n/(2+n)} & \mbox{\rm ~~; if $n\geq 2$} \\
\lambda C_2 (2K)^{n/2}     & \mbox{\rm ~~; if $n\leq 2$} 
\end{array} \right.
\EEQ

Remark: the last bound in (\ref{4:gl:4}) might be improved 
by restricting $\Omega$ to a ball around the origin and applying the 
Poincar\'e inequality. If $C_{\rm P}^{(2)} K< 1/2$, one gets 
$\int_{\Omega}|\nabla a|^2 < K/(1-K C_{\rm P}^{(2)})\, |\Omega|^{1-2/n}\,
\overline{a}^{\,2}$. 
For $n=1$, the bounds might be further sharpened with the help of the
inequality
\BEQ
\int_{\Omega} |\partial u|^p \leq c\, \eps^{-\mu(p)} 
\left( \int_{\Omega} |u|\right)^p + \eps \int_{\Omega} |\partial^2 u|^p
\EEQ
where $\mu(p) =-(p-3+1/p)/p$ and 
which can be proven for $p\geq 1$ through a slight generalization of the 
proof \cite{Frie69} of the inequality (\ref{2:gl:N}) of Lemma 2. 
Since $\mu(p)>0$ if $1\leq p<p_c=(3+\sqrt{5})/2$, the method of the proposition 
can for example be used in the $p=1$ and $p=2$ cases. 

The lower bound in (\ref{1:gl:3a}) is proved similarly. First, we set 
$k=q=1$ and $r=3$ in Lemma 1. Then $j=0$. For $1\leq n\leq 3/2$, we set $p=1$,
find $\theta=2n/3\in[2/3,1]$ and
\BEQ \label{3:gl:gn1}
\int_{\Omega} |u|^3 \leq \Gamma_1 \left( \int_{\Omega} |u|\right)^{3-2n}
\left( \int_{\Omega} |\partial u|\right)^{2n}
\EEQ 
Next, for $3/2\leq n\leq 3$, we set $p=3/2$, find $\theta=2n/(n+3)\in[2/3,1]$ 
and
\BEQ \label{3:gl:gn2}
\int_{\Omega} |u|^3 \leq \Gamma_2 \left( \int_{\Omega} |u|\right)^{3(3-n)/(n+3)}
\left( \int_{\Omega} |\partial u|^{3/2}\right)^{4n/(n+3)}
\EEQ 
Finally, for $n\geq 3$, we set $p=3$, find $\theta=2n/(2n+3)\in[2/3,1)$ and
\BEQ \label{3:gl:gn3}
\int_{\Omega} |u|^3 \leq \Gamma_3 \left( \int_{\Omega} |u|\right)^{9/(2n+3)}
\left( \int_{\Omega} |\partial u|^3 \right)^{2n/(2n+3)}
\EEQ 
Dimensional analysis shows that $\Gamma_{1,2,3}$ (which stand for $C^3$ of 
Lemma 1) are independent of $|\Omega|$. For the further analysis of
(\ref{3:gl:gn3}), the proposition above states that there is a positive
constant $K_3$ such that
\newpage \typeout{** hier ist ein Seitenvorschub !! **} 
\BEA
\int_{\Omega} |\nabla b|^3 &\leq& K_3 |\Omega|^{-3/n} \int_{\Omega} b^3
\nonumber \\
&=& K_3 |\Omega|^{-3/n} \int_{\Omega} 
\left[ \overline{b}^{\,3}+3\overline{b}(b-\overline{b})^2+(b-\overline{b})^3
\right] 
\nonumber \\
&\leq&  K_3 |\Omega|^{-3/n} \left[ |\Omega|\overline{b}^{\,3} + 
3\overline{b} \int_{\Omega} |b-\overline{b}|^2 + 
\int_{\Omega} |b-\overline{b}|^3 \right]
\nonumber \\
&\leq& 5 K_3 |\Omega|^{1-3/n}\overline{b}^{\,3}
\label{4:gl:4a}
\EEA
Now, from $\partial_t\overline{b}=-\mu\overline{b^3}$, we obtain, using first
eqns.~(\ref{3:gl:gn1},\ref{3:gl:gn2},\ref{3:gl:gn3}), then the Cauchy-Schwarz
inequality and the following consequence of H\"older's inequality
$\int_{\Omega} |\nabla b|^{3/2} \leq |\Omega|^{1/4}\left( 
\int_{\Omega} |\nabla b|^{2}\right)^{3/4}$ and finally
eqns.~(\ref{4:gl:4},\ref{4:gl:4a})
\BEA
\partial_t \overline{b} &\geq& \left\{\begin{array}{ll}
-\mu\Gamma_1 |\Omega|^{-1} \left( \int_{\Omega} b\right)^{3-2n}
\left( \int_{\Omega} |\nabla b| \right)^{2n} & 
\mbox{\rm ~~; if $1\leq n\leq 3/2$} \\
-\mu\Gamma_2 |\Omega|^{-1} \left( \int_{\Omega} b\right)^{3(3-n)/(3+n)}
\left( \int_{\Omega} |\nabla b|^{3/2} \right)^{4n/(3+n)} & 
\mbox{\rm ~~; if $3/2\leq n\leq 3$} \\
-\mu\Gamma_3 |\Omega|^{-1} \left( \int_{\Omega} b\right)^{9/(2n+3)}
\left( \int_{\Omega} |\nabla b|^{3} \right)^{2n/(2n+3)} & 
\mbox{\rm ~~; if $3\leq n$} 
\end{array} \right.
\nonumber \\
&\geq& \left\{\begin{array}{ll}
-\mu\Gamma_1 |\Omega|^{-1} \left( \int_{\Omega} b\right)^{3-2n}
\left( \int_{\Omega} |\nabla b|^2 \right)^{n} & 
\mbox{\rm ~~; if $1\leq n\leq 3/2$} \\
-\mu\Gamma_2 |\Omega|^{-1+n/(n+3)} \left( \int_{\Omega} b\right)^{3(3-n)/(3+n)}
\left( \int_{\Omega} |\nabla b|^{2} \right)^{3n/(3+n)} & 
\mbox{\rm ~~; if $3/2\leq n\leq 3$} \\
-\mu\Gamma_3 |\Omega|^{-1} \left( \int_{\Omega} b\right)^{9/(2n+3)}
\left( \int_{\Omega} |\nabla b|^{3} \right)^{2n/(2n+3)} & 
\mbox{\rm ~~; if $3\leq n$} 
\end{array} \right.
\nonumber \\
&\geq& \left\{\begin{array}{ll}
-\mu\Gamma_1 (2K)^n\, \overline{b}^{\,3} & 
\mbox{\rm ~~; if $1\leq n\leq 3/2$} \\
-\mu\Gamma_2 (2K)^{3n/(n+3)} \, \overline{b}^{\,3} & 
\mbox{\rm ~~; if $3/2\leq n\leq 3$} \\
-\mu\Gamma_3 (5K_3)^{2n/(2n+3)}\, \overline{b}^{\,3} & 
\mbox{\rm ~~; if $3\leq n$} 
\end{array} \right.
\EEA
which is the form asserted in (\ref{1:gl:3a}). We identify the effective
reaction rate
\BEQ
\mu' := \left\{\begin{array}{ll}
\mu\Gamma_1 (2K)^n & 
\mbox{\rm ~~; if $1\leq n\leq 3/2$} \\
\mu\Gamma_2 (2K)^{3n/(n+3)} & 
\mbox{\rm ~~; if $3/2\leq n\leq 3$} \\
\mu\Gamma_3 (5K_3)^{2n/(2n+3)} & 
\mbox{\rm ~~; if $3\leq n$} 
\end{array} \right.
\EEQ
This completes the proof of the theorem. \hfill q.e.d.

The bounds (\ref{1:gl:4}) are established by direct integration, since
$\partial_t \overline{a}(t) \gtrless - \lambda \overline{a}(t)^2$ implies
$\overline{a}(t) \gtrless a_0/(1+a_0\lambda t)$ from which the assertion is
immediate, under the stated condition on $t$ and for $a_0>0$. 
The bounds for $\overline{b}(t)$ follow similarly. 

We now prove the corollary. First, 
$|\Omega|^{-1}\int_{\Omega} (a-\overline{a})^2 \leq 
|\Omega|^{-1}\int_{\Omega} \overline{a}^{\,2}
=\overline{a}^{\,2}\leq (\lambda t)^{-2}$. 
Next, we try to improve this bound by letting $\Omega=B_R(0)$ be a ball around
the origin and apply Poincar\'e's inequality
\begin{displaymath}
\frac{1}{|\Omega|} \int_{\Omega} (a-\overline{a})^2 \leq
C_{\rm P}^{(2)} |\Omega|^{-1+2/n} \int_{\Omega} |\nabla a|^2 
\leq 2 K C_{\rm P}^{(2)}\, \overline{a}^{\,2}
\end{displaymath}
and therefore $K'=\min(1, 2 K C_{\rm P}^{(2)})$. 
The estimates for $b$ are obtained similarly. \hfill q.e.d. 

Remark: for the {\sc pcpd} with $D=1$, eqns.~(\ref{1:gl:10},\ref{1:gl:11}) are 
obtained by applying the results of the theorem separately to the two terms on
the right-hand side of eqn.~(\ref{1:gl:8}). 
Eq.~(\ref{1:gl:12}) is obtained by replacing $a\mapsto a- a_{\infty}$ in the 
corollary and then using (\ref{1:gl:9}). Let $a_t(\vec{r}):=a(t,\vec{r})$. 
For sufficiently long times, $a_t\in\wit{W}^{2,3}(\mathbb{R}^n)$, 
but if $\lambda>0$, $a_t\not\in W^{2,3}(\mathbb{R}^n)$.

\newpage 
\noindent {\large\bf Acknowledgments}\\

\noindent
I thank R. Cherniha, H. Spohn, M. Struwe and P. Wittwer for useful 
conversations or correspondence and the organizers of the XIV$^{\rm th}$ 
International Conference on Mathematical Physics in Lisbon, 
especially J.-C. Zambrini, for providing 
the stimulating atmosphere where the ideas presented here formed. The
warm hospitality of the Centro de F\'{\i}sica Te\'orica e Computacional 
(CFTC) of the Universidade de Lisboa is gratefully acknowledged.  


\end{document}